\documentclass[journal,twoside,web]{ieeecolor}
\usepackage{tmi}
\usepackage{cite}
\usepackage{amsmath,amssymb,amsfonts}
\usepackage{algorithmic}
\usepackage{graphicx}
\usepackage{textcomp}
\def\BibTeX{{\rm B\kern-.05em{\sc i\kern-.025em b}\kern-.08em
    T\kern-.1667em\lower.7ex\hbox{E}\kern-.125emX}}
\begin{document}
\title{Context-aware PolyUNet for Liver and Lesion Segmentation from Abdominal CT Images}
\author{Liping Zhang and Simon Chun-Ho Yu
\thanks{Liping Zhang and Simon Chun-Ho Yu are with the Department of Imaging and Interventional Radiology, Faculty of Medicine, The Chinese University of Hong Kong, Hong Kong, China (e-mail: lpzhang@link.cuhk.edu.hk; simonyu@cuhk.edu.hk).}}

\maketitle

\begin{abstract}
Accurate liver and lesion segmentation from computed tomography (CT) images are highly demanded in clinical practice for assisting the diagnosis and assessment of hepatic tumor disease.
However, automatic liver and lesion segmentation from contrast-enhanced CT volumes is extremely challenging due to the diversity in contrast, resolution, and quality of images.
Previous methods based on UNet for 2D slice-by-slice or 3D volume-by-volume segmentation either lack sufficient spatial contexts or suffer from high GPU computational cost, which limits the performance.
To tackle these issues, we propose a novel context-aware PolyUNet for accurate liver and lesion segmentation.
It jointly explores structural diversity and consecutive t-adjacent slices to enrich feature expressive power and spatial contextual information while avoiding the overload of GPU memory consumption.
In addition, we utilize zoom out/in and two-stage refinement strategy to exclude the irrelevant contexts and focus on the specific region for the fine-grained segmentation.
Our method achieved very competitive performance at the MICCAI 2017 Liver Tumor Segmentation (LiTS) Challenge among all tasks with a single model and ranked the $3^{rd}$, $12^{th}$, $2^{nd}$, and $5^{th}$ places in the liver segmentation, lesion segmentation, lesion detection, and tumor burden estimation, respectively.
\end{abstract}


\section{Introduction}
\label{sec:introduction}
\IEEEPARstart{L}{iver}
cancer is one of the most common malignant diseases worldwide and is the major cause of death from cancer \cite{bosch2004primary}.
The segmentation and 3D representation of the liver from a computed tomography (CT) scan is an important step for liver cancer diagnosis and treatment planning, which is highly demanded in clinical practice.
However, it is time-consuming for radiologists to manually delineate the liver and liver lesion slice-by-slice from a CT volume.
Besides, the manual delineation is subjective and poorly reproducible, which prone to inter- and intra-rater variations.
Therefore, automatic segmentation methods are highly demanded in clinical practice for eliminating subjectivity and cumbersome processes while producing reproducible and robust results.

Automatic liver segmentation from contrast-enhanced CT images is extremely challenging due to the low-intensity contrast between the liver and its neighboring organs as well as artifacts from injected contrast medium. It is even more difficult for the liver lesion segmentation due to the various size, shapes, and locations across different patients.
Moreover, the diversity in the contrast enhancement, resolution, and quality of CT images due to different scanners and acquisition protocols further poses challenges for automatic segmentation methods.

Many methods \cite{hann2000diagnostic, schwier2011object, heimann2009comparison} have been proposed to tackle these issues. However, these methods heavily rely on hand-crafted features, such as color, shape, and local textures, and have limited feature representation capability which results in ill-segmentation.
Recently, the fully convolutional network (FCN) \cite{long2015fully} based methods have achieved consecutive success in tackling semantic segmentation problems for natural images. It also arises a brand-new trend to use this promising technique to advance medical image analysis.
The U-Net \cite{ronneberger2015u} learns hierarchical feature representation in data-driven and end-to-end manners by combining low-level and high-level features using long-range connections, which has achieved great success in biomedical image segmentation.
Christ \textit{et al.} \cite{christ2016automatic} demonstrated the cascaded 2D U-Net and 3D conditional random fields for automatic liver and lesion segmentation from CT abdomen images. Alternatively, Dou \textit{et al.} \cite{dou20163d} exploited 3D contextual information by proposing a 3D deeply supervised network for automatic liver segmentation from CT volumes.
The above methods based on UNet for 2D slice-by-slice or 3D volume-by-volume segmentation either lack sufficient spatial contexts or suffer from high computational cost and GPU memory consumption, which limits the performance.

In this paper, we propose a novel context-aware PolyUNet for accurate liver and lesion segmentation from contrast-enhanced abdominal CT images.
The PolyUNet explores structural diversity beyond depth and width by integrating multiple paths in a “polynomial” combination fashion.
Besides, the UNet long-range skip connections \cite{ronneberger2015u} is adopted in our network to jointly combine the low-level features and its high-level counterparts for multi-scale context exploration and high-resolution feature preservation.
These architecture designs enhance the network expressive power for handling various size, shape, and locations of the liver and lesion during prediction.
Furthermore, our method leverages consecutive t-adjacent slices to jointly capture intra- and inter-slice features for accurate segmentation.
This context-aware method allows the network to effectively obtain 3D spatial contexts in a 2D manner while avoiding the overload of computational cost and GPU memory consumption as in the 3D network \cite{dou20163d}.
In addition, we jointly adopt zoom out/in approach and two-stage refinement strategy to exclude the irrelevant surrounding contexts from the intra-slice and fine-tune the candidate region for the fine-grained liver and lesion segmentation.
Overall, the major contributions of this work consist in several aspects:
\begin{enumerate}
	\item We propose a novel PolyUNet for liver and lesion segmentation, where the “polynomial” combination of multiple paths and UNet long-range skip connections are integrated to empower the network toward accurate performance.
	\item We explore a context-aware method by leveraging consecutive t-adjacent slices to jointly capture intra- and inter-slice features, which enriches 3D spatial contexts in the 2D manner while avoiding the overload of computational cost and GPU memory consumption.
	\item We adopt the two-stage refinement strategy and integrate the zoom out/in approach to exclude the irrelevant surrounding contexts and fine-turn the candidate region for fine-grained segmentation.
	\item Our method achieved very competitive performance at the LiTS-MICCAI 2017 Challenge with a single model and ranked the $3^{rd}$, $12^{th}$, $2^{nd}$, and $5^{th}$ places in the liver segmentation, lesion segmentation, lesion detection, and tumor burden estimation, respectively.
\end{enumerate}

\section{Methodology}
\begin{figure*}
	\centering
	\includegraphics[width=\textwidth]{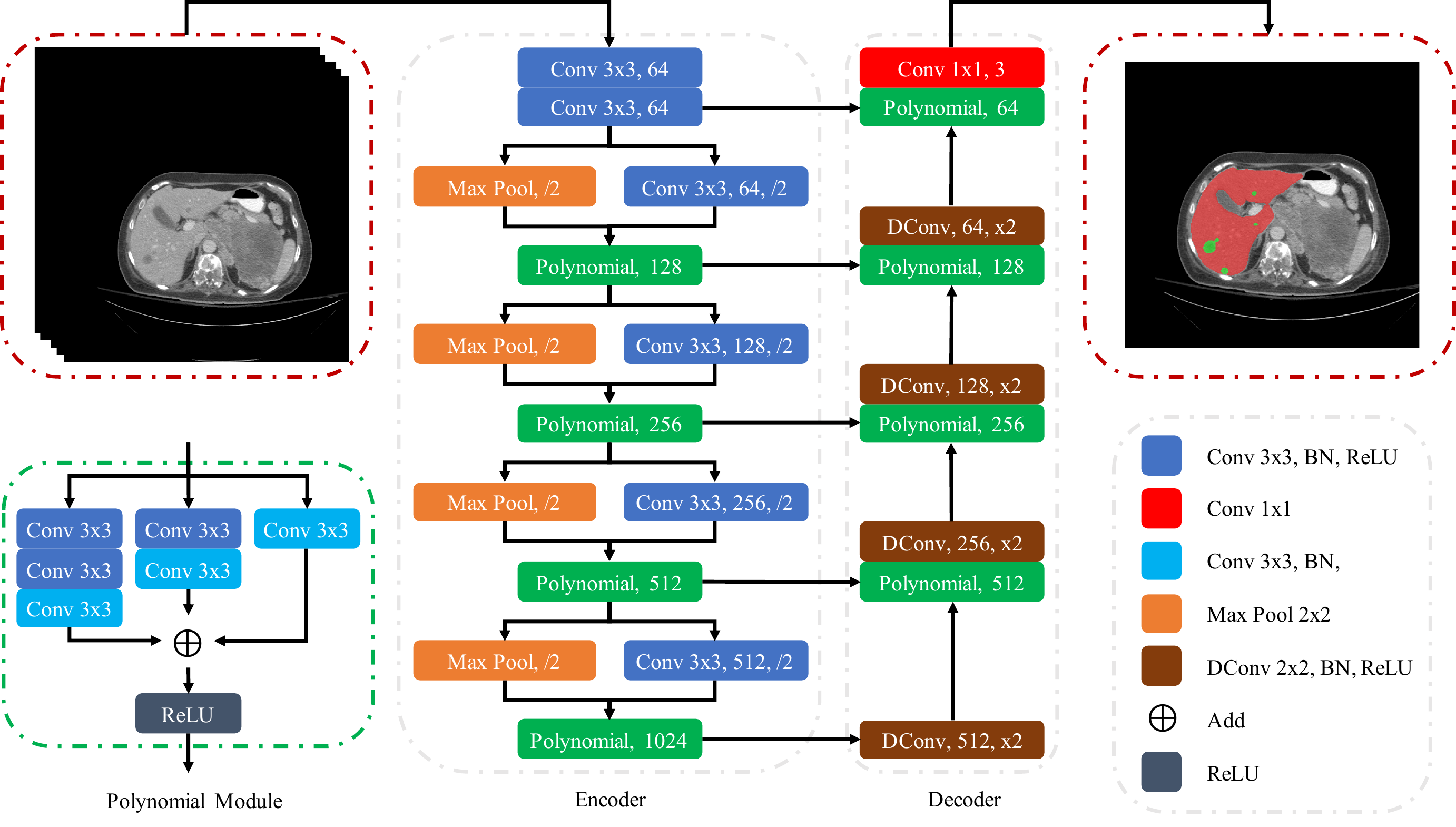}
	\caption{
		Overview architecture of the context-aware PolyUNet.
		The “x2” represents upsampling with factor 2.
		The “/2” indicates downsampling with factor 2.
	}
	\label{fig:polyunet}
\end{figure*}

\subsection{Context-aware PolyUNet}
The proposed context-aware PolyUNet architecture is illustrated in Fig.~\ref{fig:polyunet}.
The PolyUNet belongs to the 2D FCN \cite{long2015fully} family in which takes an input of the arbitrary size and produces the correspondingly-sized output with efficient inference and learning.
Specifically, it takes a set of consecutive slices as the input and learns effective and representative features through five encoding stages.
The first stage of the encoder consists of two convolution layers with $3\times3$ kernel size for low-level but high resolution features learning.
Starting from the second encoding stage, the down-sampling operation is applied first to reduce the spatial size of features for learning abstract and robust feature representation for liver and lesion.
To alleviate the spatial information loss caused by the max-pooling, we expand filter banks by concatenating additional convolution features to avoid the representational bottleneck during feature compression \cite{szegedy2016rethinking}.
The compressed features then feed into the polynomial module to learn more expressive and powerful features by integrating multiple paths in a "polynomial" combination fashion.
The rest of the encoding stages have a similar structure design like stage two except for the varied feature numbers.
There are five decoding stages in the PolyUNet to gradually refine features for high-resolution liver and lesion segmentation.
The U-Net \cite{ronneberger2015u} long-range skip connection is utilized to concatenate the high-level feature and its corresponding low-level counterpart for preserving fine-grained details before entering the polynomial module for expressive feature extraction.
The $1\times1$ convolution with the channel dimension of 3 is employed to generate a probability map for liver and lesion prediction.
It is worth note that all $3\times3$ convolution and $2\times2$ deconvolution layers are followed with the batch normalization (BN) \cite{ioffe2015batch} and the Rectified Linear Unit (ReLU) unless otherwise specified.

\subsection{Polynomial Module}
Increasing the depth or width of networks to empower the model capacity for pursuing optimal performance has been effectively demonstrated in previous work \cite{he2016deep, zagoruyko2016wide}.
On the other hand, the significance of increasing diversity in network topological has been widely studied in \cite{szegedy2016rethinking, huang2017densely}. Recently, PolyNet \cite{zhang2017polynet} explores structural diversity in designing deep networks, a new dimension beyond just depth and width, to improve the expressive power while preserving comparable computational cost.
Although it is originally designed for the object classification task, we inherit its advantages in network design and propose the polynomial module, as shown in the bottom left of the Fig.~\ref{fig:polyunet}, to extract diverse features of liver and lesion.
The corresponding architecture can be formulated as:
\begin{eqnarray}
	(F + F^2 + F^3) \cdot x 
	& = & F(x) + F(F(x)) \nonumber\\ && {} + F(F(F(x))),
\end{eqnarray}
where $x$ denotes the input feature, $F$ is the nonlinear transform carried out by convolution operation with BN and ReLU layers, and $F \cdot x$ indicates result of the operator $F$ acting on $x$.
This unit comprises three paths with a different number of convolution layers corresponding to the first-order, second-order, and third-order paths.
Information flows from these paths will be merged to the ReLU layer for nonlinear feature embedding.
The polynomial module allows the input signal to go through multi-level feature transformations, which can considerably increase the expressive power of the network.

\subsection{Consecutive t-adjacent Slices}
Contextual information plays a crucial role for a network to achieve accurate segmentation, especially in delineating liver and lesion from contrast-enhanced CT images.
The 2D network extracts expressive and representative features in a slice-by-slice manner for 3D volume data prediction.
This method only captures contexts within the target slice, namely intra-slice features, which can result in unsatisfactory performance in the liver and lesion segmentation due to the lack of spatial contextual information across neighboring slices, namely inter-slice features.
The 3D network takes 3D volume patches as the input which can capture both intra- and inter-slice contextual information but suffered from high computational cost and GPU memory consumption.

To capture sufficient contexts and avoid computation overload, we jointly explore the intra- and inter-slice features in a 2D manner to extract 3D spatial contextual information for better liver and lesion segmentation.
Specifically, instead of taking a single 2D image as the input, our method extracts features from consecutive t-adjacent slices where the target image and the corresponding nearest images are stacked together.
For the $k^{th}$ slice $V_{\cdot,\cdot,k}$ along the z-dimension of the volume $V$, the corresponding consecutive t-adjacent slices can be denoted as follows:
\begin{equation}
	V_{\cdot,\cdot,\{k-t, \cdots, k, \cdots, k+t\}}=[V_{\cdot,\cdot,k-t}, \cdots,V_{\cdot,\cdot,k}, \cdots, V_{\cdot,\cdot,k+t}],
\end{equation}
where the operation $[\cdot]$ refers to the concatenation of slices from the index $k-t$ to $k+t$.
This strategy allows the 2D network to jointly explore the 3D spatial contextual information through $2t+1$ consecutive slices for accurate segmentation while avoiding comparable computational costs.

\subsection{Zoom Out/In and Two-stage Refinement Strategy}
\begin{figure}
	\centering
	\includegraphics[width=\columnwidth]{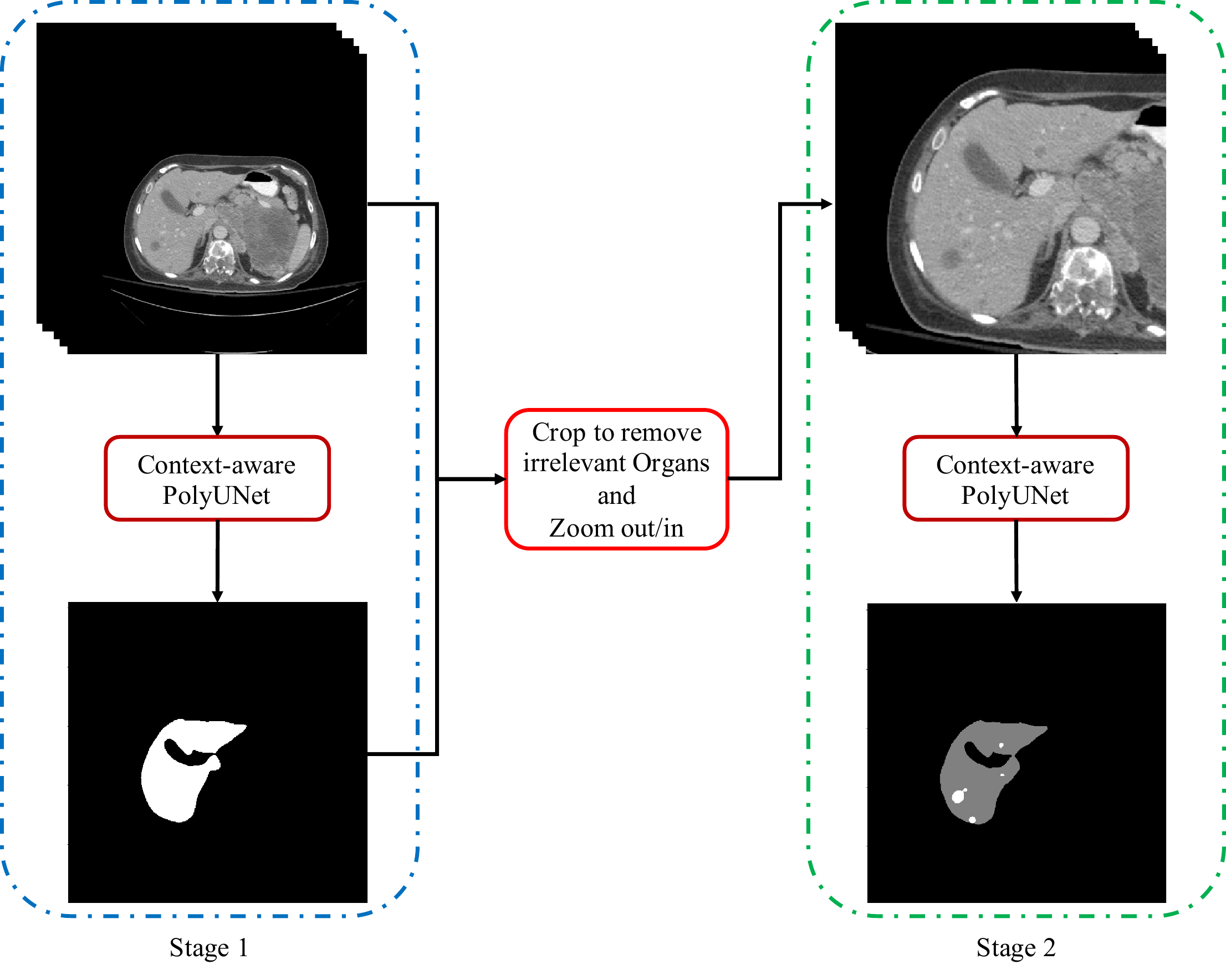}
	\caption{Our proposed zoom out/in and two-stage refinement strategy.}
	\label{fig:zoom_and_twostage}
\end{figure}

The context-aware PolyUNet extracts effective features for liver and lesion, but it may also produce over/under segmentation due to the diversity in image resolution and quality as well as the differences in size, shape and varying number of lesions.
To alleviate those problems, we integrate our network with two-stage refinement method and zoom out/in strategy to gradually obtain the fine-grained liver and lesion segmentation. The details of the framework is illustrated in Fig.~\ref{fig:zoom_and_twostage}.

The first stage takes full-resolution consecutive t-adjacent slices as the input and produces liver segmentation for the target image.
This liver mask can be used to provide the coarse region of interest (ROI) for removing irrelevant neighbor organs and tissues for the subsequent step.
However, the liver may not be entirely captured within this coarse ROI due to the under segmentation, which could cause substantial disaster for the refinement stage.
We enlarge the ROI by adding a certain padding to avoid this problem before generating the input for the second stage. Noted that, the size of padding should be chosen carefully to prevent including too much background.

The second stage intents to further refine the liver boundary and obtain lesion segmentation.
The input of this stage is generated by cropping the full-resolution consecutive t-adjacent slices using the enlarged ROI.
However, the network may still struggle from the inconsistent size of the liver and lesion across slices and scans.
Inspired by the fact that we often zoom in to a picture for observing the details of small objects and zoom out for delineating the whole picture of large objects, we proposed a very simple zoom out/in approach to transform the cropped t-adjacent slices to a certain fixed spatial resolution.
The whole strategy plays a crucial role for achieve accurate liver segmentation and lesion burden estimation.

\subsection{Training Strategy and Loss Function}
Lesions are randomly spread in the 3D liver organ and often have small size compared with liver and others.
Handling this class imbalance problem appropriately plays an important role to achieve accurate segmentation in both liver and lesions.
We employ the weighted cross-entropy as the loss function for the network training to alleviate this problem.
Given the prediction of the $k^{th}$ slice along the z-dimension from the volume $V$, the loss function for this slice can be described as follows:
\begin{equation}
	\label{eq_loss}
	\ell(Y_{\cdot,\cdot,k}, \hat{Y}_{\cdot,\cdot,k}) = -\frac{1}{|N|}
	\sum_{i=1,j=1}^N {\sum_{c=0}^C {\omega_{i,j,k}^c Y_{i,j,k}^c \log{\hat{Y}_{i,j,k}^c}}},
\end{equation}
where $N$ is the set of spatial pixel positions in the slice $V_{\cdot,\cdot,k}$, $\hat{Y}_{i,j,k}^c$, $ Y_{i,j,k}^c $, and $\omega_{i,j,k}^c$ denote the predicted probability, ground truth, and class weight of the voxel $(i,j,k)$ that belongs to the class $c$, respectively.
The class weight of the background, liver, and lesion are set to $1.0$, $2.0$, and $5.0$ in the experiments. The loss function is employed for each stage during training. Noted that, the network in the first stage only produces liver segmentation, while the network in the second stage delineates both liver and lesions.

\section{Experiments and Results}
\subsection{Datasets}
The MICCAI 2017 Liver Tumor Segmentation (LiTS) Challenge provides $201$ contrast-enhanced 3D abdominal CT scans, including $131$ for local training and $70$ for online testing.
The dataset was acquired with different CT scanners and acquisition protocols from seven clinical sites, which was considered to be very diverse with respect to image resolution and quality \cite{bilic2019liver}.
There are large variations in the in-plane image resolution from $0.56$ mm to $1.0$ mm,  inter-slice spacing from $0.45$ mm to $6.0$ mm, and the number of axial slices from $42$ to $1026$.
The liver lesions appear in $194$ CT scans, with a largely varying number from $0$ to $75$ and size from $38$ mm$^3$ to $1231$ mm$^3$.

\subsection{Evaluation Metrics}
The online evaluation system assesses both detection and segmentation for the LiTS submissions. For lesion detection, the precision and recall are calculated.
A lesion is considered detected if the predicted lesion has a greater than $50\%$ intersection over union (IOU) with the reference lesion.
For liver and lesion segmentation, the Dice score is used in two ways to assess the segmentation, with a global Dice score (combines all data sets into one) and an average Dice per volume score.
Other segmentation metrics are evaluated only for detected lesions, including volume overlap error (VOE), relative volume difference (RVD), average symmetric surface distance (ASD), maximum symmetric surface distance (MSD), and root mean square symmetric surface distance (RMSD).
If any liver prediction is not provided, all average liver segmentation metrics will default to the worst possible score (0 or infinity, depending on the metric).
The tumor burden of the liver is a measure of the fraction of the liver afflicted by cancer. We measure the root-mean-square error (RMSE) and max error in tumor burden estimation from lesion predictions.

\subsection{Implementation Details}
Networks are trained from scratch with the ``kaiming normal'' initialization manner \cite{he2015delving} under the Caffe \cite{jia2014caffe} framework. We use stochastic gradient descent (SGD) with $0.99$ momentum and weight decay $5e^{-4}$ to train our models. The initial rate is set to $1e^{-3}$ with a ``step'' learning rate policy in which the initial rate decays $0.1$ for every $40000$ iterations with total of $160000$.

For image preprocessing, we truncate the image intensity values of the CT scans to the range of $[-200, 300]$ Hounsfield Unit (HU) to exclude irrelevant organs while fully capturing liver and lesions. We also transform all scans to have zero means and unit variances. For data augmentation, the random rotation, flip, and scaling are adopted.

For post-processing, the largest connected component is performed on the foreground prediction for both stages to remove noise background.
The lesion segmentation is produced from the second stage directly, while the fine-grain liver segmentation is obtained from the combination of results of all stages.

\subsection{Results on the LiTS-MICCAI 2017 Challenge}
\begin{figure}
	\centering
	\includegraphics[width=\columnwidth]{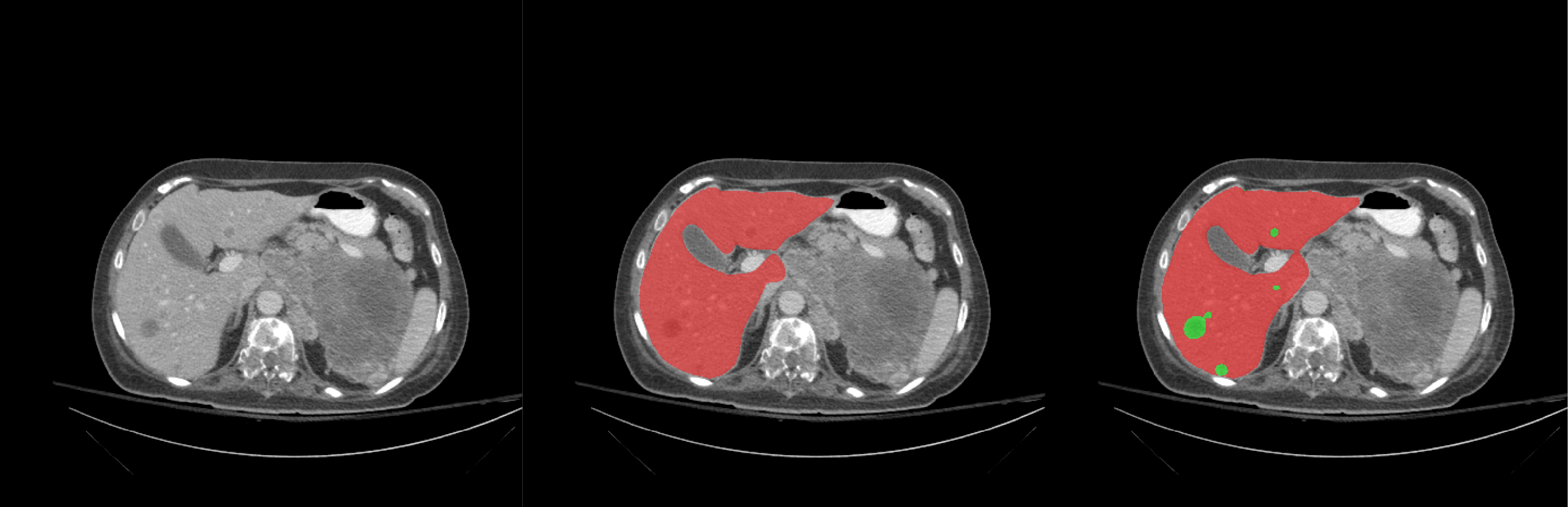}
	\caption{
		An example of liver and lesion segmentation results of our method from the test dataset. The red regions denote the liver and the green ones are the lesions.
		The three images are the input image, coarse segmentation of liver from the first stage, and fine-grained segmentation for livers and lesions.}
	\label{fig:exp1}
\end{figure}

\begin{table*}
	\centering
	\caption{Liver segmentation submissions of  LiTS-MICCAI 2017 challenge ranked by Dice per case score.}
	\begin{tabular}{c | c | c | c | c | c | c | c | c}
		\hline
		Ranking &	Name	&	Dice per case	&	Dice global	&	VOE			&	RVD			&	ASD			&	MSD			&	RMSD		\\
		\hline
		1		&	deepX 			&	0.9630 (1)	&	0.9670 (2)	&	0.071 (2)	&	-0.010 (9)	&	1.104 (1)	&	23.847 (1)	&	2.303 (1)	\\
		2		&	hchen			&	0.9610 (2)	&	0.9650 (3)	&	0.074 (3)	&	0.000 (14)	&	1.692 (9)	&	29.411 (8)	&	3.729 (10)	\\
		2		&	leHealth		&	0.9610 (2)	&	0.9640 (4)	&	0.075 (4)	&	0.023 (17)	&	1.268 (4)	&	27.016 (4)	&	2.776 (4)	\\
		3		&	mabc			&	0.9600 (3)	&	0.9700 (1)	&	0.070 (1)	&	0.000 (14)	&	1.130 (2)	&	24.450 (2)	&	2.390 (2)	\\
		3		&	hans.meine		&	0.9600 (3)	&	0.9650 (3)	&	0.077 (6)	&	-0.004 (13)	&	1.150 (3)	&	24.499 (3)	&	2.421 (3)	\\
		\textbf{3}	&	\textbf{LP777 (ours)}	&	\textbf{0.9600 (3)}	&	\textbf{0.9630 (5)}	&	\textbf{0.076 (5)}	&	\textbf{0.009 (15)}	&	\textbf{1.510 (8)}	&	\textbf{31.225 (10)}	&	\textbf{3.510 (8)}	\\
		4		&	yaya			&	0.9590 (4)	&	0.9630 (5)	&	0.079 (7)	&	-0.009 (10)	&	1.311 (5)	&	28.229 (6)	&	2.929 (5)	\\
		5		&	chunliang		&	0.9580 (5)	&	0.9620 (6)	&	0.080 (8)	&	-0.006 (12)	&	1.367 (7)	&	32.933 (14)	&	3.260 (7)	\\
		5		&	jrwin			&	0.9580 (5)	&	0.9620 (6)	&	0.081 (9)	&	-0.016 (8)	&	1.360 (6)	&	27.732 (5)	&	2.994 (6)	\\
		6		&	MICDIIR			&	0.9560 (6)	&	0.9590 (7)	&	0.083 (10)	&	0.031 (18)	&	1.847 (11)	&	35.653 (16)	&	4.393 (15)	\\
		7		&	medical			&	0.9510 (7)	&	0.9510 (10)	&	0.093 (12)	&	-0.009 (10)	&	1.785 (10)	&	29.769 (9)	&	3.933 (13)	\\
		8		&	predible		&	0.9500 (8)	&	0.9500 (11)	&	0.090 (11)	&	-0.020 (7)	&	1.880 (13)	&	32.710 (13)	&	4.200 (14)	\\
		8		&	Planckton		&	0.9500 (8)	&	0.9400 (14)	&	0.100 (13)	&	-0.050 (6)	&	1.890 (14)	&	31.930 (12)	&	3.860 (12)	\\
		9		&	huni1115		&	0.9460 (9)	&	0.9470 (12)	&	0.101 (14)	&	-0.009 (10)	&	1.869 (12)	&	31.840 (11)	&	3.788 (11)	\\
		10		&	ed10b047		&	0.9400 (10)	&	0.9400 (14)	&	0.110 (15)	&	0.050 (19)	&	2.890 (17)	&	51.220 (18)	&	7.210 (17)	\\
		10		&	bratta			&	0.9400 (10)	&	0.9400 (14)	&	0.120 (17)	&	0.060 (20)	&	3.540 (20)	&	186.250 (25)&	11.240 (23)	\\
		11		&	mbb	14			&	0.9380 (11)	&	0.9520 (9)	&	0.113 (16)	&	-0.006 (12)	&	2.900 (18)	&	90.245 (21)	&	7.902 (19)	\\
		12		&	lei	11			&	0.9340 (12)	&	0.9580 (8)	&	0.101 (14)	&	257.163 (23)&	258.598 (27)&	321.710 (26)&	261.866 (27)\\
		13		&	Micro			&	0.9320 (13)	&	0.9410 (13)	&	0.126 (18)	&	0.088 (21)	&	2.182 (16)	&	33.588 (15)	&	4.501 (16)	\\
		14		&	szm0219			&	0.9300 (14)	&	0.9410 (13)	&	0.126 (18)	&	-0.068 (4)	&	3.974 (21)	&	61.894 (19)	&	9.136 (20)	\\
		14		&	MIP\_HQU		&	0.9300 (14)	&	0.9400 (14)	&	0.120 (17)	&	0.010 (16)	&	2.160 (15)	&	29.270 (7)	&	3.690 (9)	\\
		15		&	jinqi			&	0.9240 (15)	&	0.9230 (17)	&	0.140 (19)	&	-0.052 (5)	&	5.104 (23)	&	123.332 (23)&	13.464 (24)	\\
		16		&	mahendrakhened	&	0.9120 (16)	&	0.9230 (17)	&	0.150 (20)	&	-0.008 (11)	&	6.465 (24)	&	45.928 (17)	&	9.682 (21)	\\
		17		&	zachary			&	0.9100 (17)	&	0.9300 (16)	&	0.150 (20)	&	-0.070 (3)	&	4.380 (22)	&	126.290 (24)&	11.220 (22)	\\
		18		&	jkan			&	0.9060 (18)	&	0.9390 (15)	&	0.157 (21)	&	-0.107 (2)	&	3.367 (19)	&	63.232 (20)	&	7.847 (18)	\\
		19		&	mapi			&	0.7670 (19)	&	0.7790 (18)	&	0.371 (22)	&	0.606 (22)	&	37.450 (26)	&	326.334 (27)&	70.879 (26)	\\
		20		&	QiaoTian		&	0.0500 (20)	&	0.0600 (20)	&	0.970 (23)	&	-0.970 (1)	&	31.560 (25)	&	90.470 (22)	&	36.800 (25)	\\
		21		&	Njust768		&	0.0410 (21)	&	0.1350 (19)	&	0.973 (24)	&	8229.525 (24)&	8231.318 (28)&	8240.644 (28)&	8232.225 (28)\\
		\hline
	\end{tabular}
	\label{tab:liver_segmentation}
\end{table*}

\begin{table*}
	\centering
	\caption{Lesion segmentation submissions of LiTS-MICCAI 2017 challenge ranked by Dice per case score.}
	\begin{tabular}{c | c | c | c | c | c | c | c | c}
		\hline
		Ranking &	Name	&	Dice per case	&	Dice global	&	VOE			&	RVD			&	ASD			&	MSD			&	RMSD		\\
		\hline
		1		&	leHealth		&	0.7020 (1)	&	0.7940 (5)	&	0.394 (11)	&	5.921 (18)	&	1.189 (12)	&	6.682 (5)	&	1.726 (8)	\\
		2		&	hchen			&	0.6860 (2)	&	0.8290 (1)	&	0.356 (3)	&	5.164 (17)	&	1.073 (5)	&	6.055 (1)	&	1.562 (2)	\\
		3		&	hans.meine 		&	0.6760 (3)	&	0.7960 (4)	&	0.383 (10)	&	0.464 (12)	&	1.143 (8)	&	7.322 (12)	&	1.728 (9)	\\
		4		&	medical			&	0.6610 (4)	&	0.7830 (9)	&	0.357 (4)	&	12.124 (24)	&	1.075 (6)	&	6.317 (3)	&	1.596 (3)	\\
		5		&	deepX			&	0.6570 (5)	&	0.8200 (2)	&	0.378 (9)	&	0.288 (11)	&	1.151 (9)	&	6.269 (2)	&	1.678 (5)	\\
		6		&	Njust768 		&	0.6550 (6)	&	0.7680 (12)	&	0.451 (19)	&	5.949 (19)	&	1.607 (24)	&	9.363 (25)	&	2.313 (24)	\\
		7		&	lei 			&	0.6450 (7)	&	0.7350 (16)	&	0.356 (3)	&	3.431 (13)	&	1.006 (2)	&	6.472 (4)	&	1.520 (1)	\\
		8		&	predible 		&	0.6400 (8)	&	0.7700 (11)	&	0.340 (1)	&	0.190 (9)	&	1.040 (3)	&	7.250 (11)	&	1.680 (6)	\\	
		9		&	ed10b047 		&	0.6300 (9)	&	0.7700 (11)	&	0.350 (2)	&	0.170 (8)	&	1.050 (4)	&	7.210 (9)	&	1.690 (7)	\\	
		10		&	chunliang 		&	0.6250 (10)	&	0.7880 (7)	&	0.378 (9)	&	8.300 (21)	&	1.260 (15)	&	6.983 (8)	&	1.865 (13)	\\
		11		&	yaya 			&	0.6240 (11)	&	0.7920 (6)	&	0.394 (11)	&	4.679 (14)	&	1.232 (14)	&	7.783 (17)	&	1.889 (14)	\\
		12		&	mabc 			&	0.6200 (12)	&	0.8000 (3)	&	0.360 (5)	&	0.200 (10)	&	1.290 (16)	&	8.060 (19)	&	2.000 (16)	\\
		\textbf{12}	&	\textbf{LP777 (ours)}	&	\textbf{0.6200 (12)}	&	\textbf{0.8000 (3)}	&	\textbf{0.421 (14)}	&	\textbf{6.420 (20)}	&	\textbf{1.388 (20)}	&	\textbf{6.716 (6)}	&	\textbf{1.936 (15)}	\\
		13		&	Micro 			&	0.6130 (13)	&	0.7830 (9)	&	0.430 (16)	&	5.045 (16)	&	1.759 (27)	&	10.087 (26)	&	2.556 (25)	\\	
		13		&	jrwin 			&	0.6130 (13)	&	0.7640 (13)	&	0.361 (6)	&	4.993 (15)	&	1.164 (11)	&	7.649 (15)	&	1.831 (12)	\\
		14		&	mbb 			&	0.5860 (14)	&	0.7410 (15)	&	0.429 (15)	&	39.763 (27)	&	1.649 (26)	&	8.079 (20)	&	2.252 (23)	\\
		15		&	szm0219 		&	0.5850 (15)	&	0.7450 (14)	&	0.364 (7)	&	0.001 (4)	&	1.222 (13)	&	7.408 (13)	&	1.758 (10)	\\
		16		&	MICDIIR 		&	0.5820 (16)	&	0.7760 (10)	&	0.446 (18)	&	8.775 (22)	&	1.588 (23)	&	7.723 (16)	&	2.182 (22)	\\
		17		&	Planckton 		&	0.5700 (17)	&	0.6600 (20)	&	0.340 (1)	&	0.020 (5)	&	0.950 (1)	&	6.810 (7)	&	1.600 (4)	\\
		18		&	jkan 			&	0.5670 (18)	&	0.7840 (8)	&	0.364 (7)	&	0.112 (7)	&	1.159 (10)	&	7.230 (10)	&	1.690 (7)	\\
		19		&	zachary 		&	0.5000 (19)	&	0.7200 (17)	&	0.360 (5)	&	-0.010 (3)	&	1.340 (18)	&	8.740 (23)	&	2.050 (18)	\\
		20		&	huni1115 		&	0.4960 (20)	&	0.7000 (18)	&	0.400 (12)	&	0.060 (6)	&	1.342 (19)	&	9.030 (24)	&	2.041 (17)	\\
		21		&	mahendrakhened 	&	0.4920 (21)	&	0.6250 (23)	&	0.411 (13)	&	19.705 (26)	&	1.441 (21)	&	7.515 (14)	&	2.070 (19)	\\
		22		&	bratta 			&	0.4800 (22)	&	0.7000 (18)	&	0.360 (5)	&	0.060 (6)	&	1.330 (17)	&	8.640 (22)	&	2.100 (20)	\\
		23		&	jinqi 			&	0.4710 (23)	&	0.6470 (22)	&	0.514 (20)	&	17.832 (25)	&	2.465 (28)	&	14.588 (28)	&	3.643 (27)	\\
		24		&	MIP\_HQU		&	0.4700 (24)	&	0.6500 (21)	&	0.340 (1)	&	-0.130 (1)	&	1.090 (7)	&	7.840 (18)	&	1.800 (11)	\\
		25		&	mapi 			&	0.4450 (25)	&	0.6960 (19)	&	0.445 (17)	&	10.121 (23)	&	1.464 (22)	&	8.391 (21)	&	2.136 (21)	\\
		26		&	QiaoTian 		&	0.2500 (26)	&	0.4500 (24)	&	0.370 (8)	&	-0.100 (2)	&	1.620 (25)	&	11.720 (27)	&	2.620 (26)	\\
		\hline
	\end{tabular}
	\label{tab:lesion_segmentation}
\end{table*}

\begin{table}
	\centering
	\caption{Lesion-level precision and recall scores in LiTS-MICCAI 2017 challenge.}
	\begin{tabular}{c | c | c | c}
		\hline
		Ranking &	Name			&	Precision	&	Recall	\\
		\hline
		1		&	MICDIIR			&	0.143 (15)	&	0.463 (1)	\\
		\textbf{2}	&	\textbf{LP777 (ours)}	&	\textbf{0.239 (9)}	&	\textbf{0.446 (2)}	\\
		3		&	leHealth		&	0.156 (14)	&	0.437 (3)	\\
		4		&	hchen			&	0.409 (4)	&	0.408 (4)	\\
		5		&	deepX			&	0.328 (5)	&	0.397 (5)	\\
		5		&	hans.meine 		&	0.496 (2)	&	0.397 (5)	\\
		6		&	medical			&	0.446 (3)	&	0.374 (6)	\\
		7		&	yaya			&	0.179 (12)	&	0.372 (7)	\\
		8		&	mbb				&	0.054 (23)	&	0.369 (8)	\\
		9		&	chunliang		&	0.160 (13)	&	0.349 (9)	\\
		10		&	mahendrakhened	&	0.117 (17)	&	0.348 (10)	\\
		11		&	lei				&	0.315 (6)	&	0.343 (11)	\\
		12		&	predible		&	0.140 (16)	&	0.330 (12)	\\
		12		&	ed10b047		&	0.160 (13)	&	0.330 (12)	\\
		13		&	Micro			&	0.095 (19)	&	0.328 (13)	\\
		14		&	mapi			&	0.068 (21)	&	0.325 (14)	\\
		15		&	Planckton		&	0.070 (20)	&	0.300 (15)	\\
		16		&	mabc			&	0.270 (7)	&	0.290 (16)	\\
		16		&	jrwin			&	0.241 (8)	&	0.290 (16)	\\
		17		&	Njust768		&	0.499 (1)	&	0.289 (17)	\\
		18		&	jkan			&	0.218 (11)	&	0.250 (18)	\\
		19		&	szm0219			&	0.224 (10)	&	0.239 (19)	\\
		20		&	jinqi			&	0.044 (24)	&	0.232 (20)	\\
		21		&	MIP\_HQU		&	0.030 (26)	&	0.220 (21)	\\
		22		&	huni1115		&	0.041 (25)	&	0.196 (22)	\\
		23		&	bratta			&	0.060 (22)	&	0.190 (23)	\\
		24		&	zachary			&	0.100 (18)	&	0.180 (24)	\\
		25		&	QiaoTian		&	0.010 (27)	&	0.060 (25)	\\
		\hline
	\end{tabular}
	\label{tab:lesion_detection}
\end{table}

\begin{table}
	\centering
	\caption{Tumor burden ranking in LiTS-MICCAI 2017 challenge.}
	\begin{tabular}{c | c | c | c}
		\hline
		Ranking &	Name			&	RMSE		&	Max Error	\\
		\hline
		1		&	hchen			&	0.0150 (1)	&	0.0620 (8)	\\
		2		&	yaya 			&	0.0160 (2)	&	0.0480 (4)	\\
		2		&	chunliang 		&	0.0160 (2)	&	0.0580 (6)	\\
		3		&	deepX			&	0.0170 (3)	&	0.0490 (5)	\\
		3		&	leHealth		&	0.0170 (3)	&	0.0450 (3)	\\
		4		&	predible		&	0.0200 (4)	&	0.0900 (14)	\\
		4		&	ed10b047		&	0.0200 (4)	&	0.0800 (12)	\\
		4		&	mabc			&	0.0200 (4)	&	0.0700 (10)	\\
		4		&	jrwin			&	0.0200 (4)	&	0.0860 (13)	\\
		4		&	hans.meine 		&	0.0200 (4)	&	0.0700 (10)	\\
		\textbf{5}	&	\textbf{LP777 (ours)}	&	\textbf{0.0220 (5)}	&	\textbf{0.0740 (11)	}\\
		6		&	medical 		&	0.0230 (6)	&	0.1120 (17)	\\
		6		&	jkan			&	0.0230 (6)	&	0.0680 (9)	\\
		6		&	szm0219			&	0.0230 (6)	&	0.0940 (16)	\\
		7		&	huni1115		&	0.0260 (7)	&	0.1160 (18)	\\
		7		&	MICDIIR			&	0.0260 (7)	&	0.0450 (3)	\\
		8		&	Micro			&	0.0270 (8)	&	0.0610 (7)	\\
		9		&	zachary			&	0.0300 (9)	&	0.1800 (21)	\\
		9		&	bratta			&	0.0300 (9)	&	0.1400 (19)	\\
		9		&	Planckton		&	0.0300 (9)	&	0.1800 (21)	\\
		10		&	mapi			&	0.0370 (10)	&	0.1430 (20)	\\
		11		&	MIP\_HQU		&	0.0400 (11)	&	0.1400 (19)	\\
		12		&	jinqi			&	0.0420 (12)	&	0.0330 (2)	\\
		13		&	mahendrakhened	&	0.0440 (13)	&	0.1940 (22)	\\
		14		&	mbb				&	0.0540 (14)	&	0.0920 (15)	\\
		15		&	lei				&	0.1700 (15)	&	0.0740 (11)	\\
		16		&	Njust768		&	0.9200 (16)	&	0.0610 (7)	\\
		17		&	QiaoTian		&	0.9500 (17)	&	-0.6500 (1)	\\
		\hline
	\end{tabular}
	\label{tab:tumor_burden}
\end{table}

The LiTS challenge received $61$ valid submissions in 2017 ISBI and 2017 MICCAI.
The same training and test datasets are employed in both challenges for fair comparison.
More metrics have been evaluated in the 2017 MICCAI challenge for comprehensive comparison.
Our team participated the LiTS-MICCAI 2017 Challenge and achieved very competitive performance.
The details of submissions can be found in the challenge evaluation website \footnote[1]{https://competitions.codalab.org/competitions/17094\#results} and also be reported in \cite{bilic2019liver}.

The liver segmentation results are shown in Table~\ref{tab:liver_segmentation}.
It can be observed that our proposed method achieved the state-of-the-art results with the $0.9600$ Dice per case score and ranked the $3^{rd}$ place among all the teams.
For lesion evaluation, both the segmentation and detection performance are evaluated, as they are closely linked. Their results are shown in Table~\ref{tab:lesion_segmentation} and Table~\ref{tab:lesion_detection}, respectively.
We achieved $0.6200$ Dice per case score for the segmentation of lesions with the ranking of $12^{th}$. On the other hand, we achieved a very high recall with $0.446$ for detecting lesions and ranked the $2^{nd}$ place among all the teams.
The tumor burden of the liver is a measure of the fraction of the liver afflicted by cancer.
In this task, we achieved very competitive performance with $0.022$ RMSE and ranked the $5^{th}$ place among all the teams, as shown in Table~\ref{tab:tumor_burden}.

Overall, we achieved very competitive performance in all tasks of the submissions of  LiTS-MICCAI 2017 challenge using only a single model, which demonstrates the effectiveness of our proposed method.
A representative example from the test dataset is shown in Fig.~\ref{fig:exp1}.

\section{Conclusion}
We proposed context-aware PolyUNet for liver and lesion segmentation from contrast-enhanced Abdominal CT images, which explores structural diversity to enhance the feature expressive power by integrating multiple paths in a “polynomial” combination fashion.
The network leverages consecutive t-adjacent slices to jointly capture intra-slice and inter-slice contexts while preventing from heavy computational costs.
We further adopted the zoom out/in and two-stage refinement strategy to exclude the irrelevant contexts and gradually refine the specific region for the fine-grained segmentation. 
The zoom out/in and two-stage refinement strategy further increases the segmentation performance by exclude the irrelevant contexts and focusing on the specific region.
With a single model, our method achieved superior results on the LiTS-MICCAI 2017 challenge and ranked $3^{rd}$, $12^{th}$, $2^{nd}$, and $5^{th}$ in liver segmentation, lesion segmentation, lesion detection, and tumor burden estimation, respectively.

\bibliographystyle{IEEEtran}
\bibliography{main}

\end{document}